\documentclass[square]{ws-procs975x65}
\usepackage{graphicx}
\begin{document}

\title{LEOPOLD ERNST HALPERN AND THE\\GENERALIZATION
   OF GENERAL RELATIVITY}

\author{JAMES M. OVERDUIN}

\address{Gravity Probe B, Hansen Experimental Physics Laboratory, Stanford
   University, Stanford, California 94305, U.S.A.\\
   E-mail: overduin@relgyro.stanford.edu}

\author{HANS S. PLENDL}

\address{Department of Physics, Florida State University, Tallahassee,
   Florida 32301, U.S.A.\\
   E-mail: Plendl@phy.fsu.edu}

\begin{abstract}
Leopold Ernst Halpern, who was a close associate of both Erwin Schr\"odinger
and Paul Dirac before making his own mark as a theoretical physicist of the
first rank, died in Tallahassee, Florida on 3 June 2006 after a valiant
struggle with cancer.  We give an outline of his life and work, including
his progress towards a unified gauge theory of gravitation and spin.
\end{abstract}


\bodymatter

\section*{Leopold Ernst Halpern}

Leopold was born in 1925 in Vienna, Austria.  When Hitler's armies invaded
that country in 1938, he and his parents managed to get on the last available
boat to British-controlled Palestine, where the family eked out an existence
in the desert near Tel Aviv.  Here Leopold attended a public high school,
learned Arabic and discovered his love of physics, a development that did not
sit well with his parents, who had left everything behind in Austria and told
him, ``We cannot afford a scientist!''  He was stubborn as well as idealistic,
as shown by a story that he refused to ride in the armored schoolbus on field
trips, insisting instead to be allowed to follow the bus on his bicycle so
that he could mingle freely with the local population.  When the war ended,
he returned to Vienna to become a scientist, a decision that estranged him
permanently from his parents, who wanted him to remain behind with them and
enter the family accounting business.  In his first years in Austria he
devoted much of his spare time to helping concentration-camp survivors and
returning prisoners of war.  Most of his own extended family had been killed
in the Holocaust.

Leopold's scientific career began in experimental solid-state physics.
He completed his doctorate at the University of Vienna in 1952 with a thesis
on magnetoresistance in bismuth [1-4] and took up a position at Rensselaer
Polytechnic Institute in Troy, New York as a Fulbright Scholar (1952-3).
Growing dissatisfied with purely experimental work, however, he returned
to Vienna to teach himself theoretical physics, concentrating on general
relativity (a subject not then taught in Austria).  His rapidly growing
expertise in this field attracted the attention of 1933 Nobel Laureate
Erwin Schr\"odinger, who invited Leopold to join him as an assistant until
his own retirement from active research (1956-9).  The two worked together on
several projects during this period, none of which stamped Leopold more
indelibly than the problem of understanding why gravity is so weak.
His response to this challenge was to mount a systematic and unprecedented
exploration of the frontier between gravitation and the quantum world,
looking for ways in which gravity might have some observable quantum effect.
This search took him through a succession of research positions at CERN
(1959-60), the Institute of Field Physics at the University of North Carolina,
Chapel Hill (1960-1), the Niels Bohr Institute in Copenhagen (1962-3), the
Institute for Theoretical Physics at the University of Stockholm (1963-6),
the Institut Henri Poincar\'e in Paris (1966-7), the University of Windsor
in Canada (1967-70), the Institute of Fundamental Physics in Kyoto (1970),
the Universit\'e Libre de Bruxelles (1970-3), the International Centre for
Theoretical Physics in Trieste (1973) and the Institute for Theoretical
Physics at the University of Amsterdam (1973-4).  He investigated such
ground-breaking topics as the interaction between classical gravitational
fields and quantized matter fields, exotic elementary-particle processes
(e.g. the transition of a photon into three photons, gravitational radiation
of photons, photon pair creation by gravitational radiation), the sources,
detection and intensity of gravitational radiation in the Universe, the
possibility of stimulated photon-graviton conversion by electromagnetic
fields and its corollary, the gravitational laser or ``gaser'' [5-34].
In each case, he found that the quantum face of gravity retreated behind a
veil of secrecy, largely because the desired signal would be drowned out by
noise from competing non-gravitational processes.  Leopold's calculations
suggested, for instance, that a suitable gaser would need to stretch from
the Earth to the Moon or beyond, and that its construction cost would exceed
the combined military budgets of both superpowers -- though (he wrote in 1987)
it is ``questionable whether they would agree to invest in that direction to
experience the pleasure of intense gravitational radiation before mutual
destruction''\cite{leo87c}.  Eventually (as he told us in 2004), he concluded
that gravity would never reveal itself in this way, and even began to doubt
whether it could be quantized at all.  The time was ripe for a new approach
to the problem of unification.

The spark for this new approach came in 1974 when Leopold received an offer
to take up a position at Florida State University in Tallahassee with the
other winner of the 1933 Nobel Prize in physics, Paul Adrien Maurice Dirac.
Dirac was to exert an influence on Leopold's career as strong as that of
Schr\"odinger.  Within one year he turned away from the frontier between
general relativity and quantum theory, and instead began constructing a new
gauge theory of gravitation from scratch: one that would reduce to Einstein's
theory in the appropriate limits, but would also provide a geometrical basis
for quantum phenomena such as particle spin.  The inspiration for this theory
came from Dirac's demonstration that the equation of motion for spinning
particles (the famous Dirac equation) could be expressed in the language
of group theory (specifically, in terms of the generators of the
de~Sitter group).  This demonstration suggested to Leopold
a way to generalize general relativity by building spin explicitly into the
principle of inertia.  Paraphrasing Newton, and later Einstein, he summed up
his new version of this principle as follows: ``A particle --- structureless
or spinning --- moves along the projection of an orbit of the de~Sitter group
on the de~Sitter universe, unless acted on by external forces.''  The unified
theory of spin and gravitation that grew out of this idea would occupy him
for the rest of his career [35-87].  It would take us too far afield to
review this work in detail, but we note for interested readers that Leopold
expressed the view in early 2006 that the best exposition of his theory was
to be found in Ref.~82.  Nearly all this research was carried out in
Tallahassee, with the exception of visiting positions at the Stanford Linear
Accelerator Center in California (1978) and the Universities of Amsterdam
(1983) and Stockholm (1985).  The potential significance of Leopold's ideas
can be judged by a story from the 1980s, when his visa expired and he 
received a letter from U.S. immigration authorities warning of possible
deportation.  A friend who heard of this reported the situation to Nobel
Laureate Eugene Wigner, who immediately telephoned the President of the U.S.A.
and informed him that unless Leopold received permission to stay in the
country, ``national science will be set back for years.''  Leopold received
word that his papers were on the way within hours from the head of the
Immigration and Naturalization Service himself.

When asked in 2005 whether his theory had given him any ``happiest moments''
akin to Einstein's realization that he could explain the perihelion shift
of Mercury, Leopold answered that it had.  The greatest of these was the
realization that the mathematics of the theory required both ``inner'' and
``outer manifestations'' of dynamical variables --- i.e., it naturally
accounted for both spin and angular momentum (technically, these arise from
the existence of left and right invariant vectors of the de~Sitter group).
A close second was the discovery that the field equations of the theory, when
expressed in the form of Einstein's gravitational field equations (with
extra terms), demanded the presence of a positive cosmological constant ---
i.e., vacuum energy.  Leopold regarded the observational detection of dark
energy by supernova observers in 1998 as important support for his ideas.
He also felt that his theory came closer than Einstein's to embodying
Mach's principle, in the sense that one could eliminate ``sourceless''
gravitational fields, because his theory supplied additional source terms in
the form of spin currents associated with a Yang-Mills-type gauge field
\cite{leo04a}.  Finally, influenced by Schr\"odinger as well as Einstein,
Leopold hoped that his theory would enable one to exclude the singular
solutions that plague general relativity\cite{leo04b}.

After Dirac's death in 1984, Leopold moved to the Jet Propulsion Laboratory
in Pasadena (1986-8), and then back to the Florida State University in
Tallahassee, where he remained until 2004.  These years brought him into
closer contact with the experimental gravitation community, and in particular
with a proposal to test general relativity using ultra-precise gyroscopes
in low-earth orbit.  Known as Gravity Probe~B (GPB), this experiment had its
origins in discussions between Bob Cannon, William Fairbank, Leonard Schiff
and others at Stanford University.  (Fig.~1 shows Leopold at the first
William Fairbank meeting in Rome, 1990).
\begin{figure}[t!]
\begin{center}
\includegraphics[width=110mm]{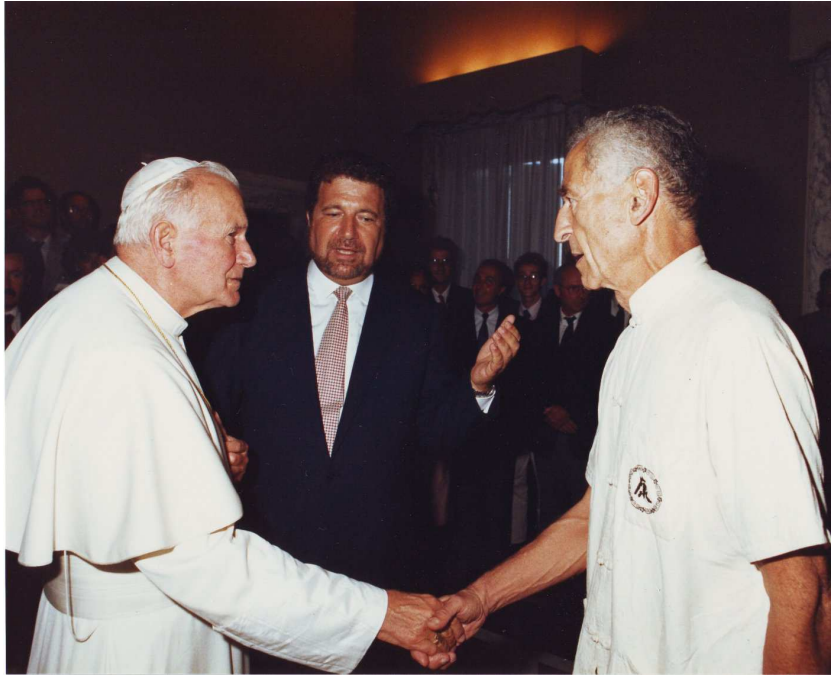}
\end{center}
\caption{Leopold Halpern (right) with Remo Ruffini and Pope John Paul II at
   the Vatican Observatory during the first William Fairbank meeting in Rome,
   1990} 
\label{fig1}
\end{figure}
Leopold joined GPB as a long-term visitor in 2004 and was an
active member of its theory group until the time of his death.  This period
overlapped with the flight of the experiment itself.  The data acquired
during this mission is now being analyzed, and will soon provide an answer
as to whether general relativity correctly predicts the behavior of spinning
test bodies in the gravitational field of the spinning Earth.
The GPB experiment will likely also provide the best test for
decades to come of extended versions of Einstein's theory like that conceived
by Leopold.  Unfortunately, the rapid progress of Leopold's cancer during his
last years prevented him from refining his theory to the point where it could
produce an unambiguous numerical prediction for the spin-axis precession of
a real test body.  The fear that he would leave his life's work incomplete
was a source of far greater anguish to Leopold than the prospect of death
itself.  He continued to wrestle with the details of his theory until his
last day.  A notepad on the desk in his office at GPB contains a final page
of group-theory calculations ending with the question, ``How to generalize?''

Leopold's interests extended far beyond physics.  He was the author of fine
historical and biographical studies, not only of his mentors Schr\"odinger
and Dirac [88-98] but also of his Dutch colleague Siegfried
Wouthuysen\cite{leo97b} and the Austrian pioneer of radioactivity research,
Marietta Blau [100-104].  It was the latter work which resulted in Leopold's
nomination to Fellowship in the American Physical Society in 2003.

A passionate environmentalist throughout his life, he was actively
engaged locally and globally in preserving the environment and wildlife for
future generations.  He had a special love for the plains of eastern Africa,
where he engaged in such activities as camping beside watering holes for
extended periods to monitor the cheetah population.  (Nobody knows exactly
how many languages he spoke fluently, but they included Swahili.)  He fought
to preserve the tropical rainforests of Sarawak in Malaysia.  In Florida,
he became especially attached to the wilderness around the Wakulla river,
and introduced many of his colleagues to its wonders.
He took pride in his physical condition, and remained active as
a swimmer, hiker and climber until his last days.  Leopold was a skilled
naturopath and traditional healer, becoming the only non-African to be
inducted into the African Medicine Man's Society in honor of the medical
care he provided there.  He participated in humanitarian projects, such as
providing food and medicine to the needy, wherever he lived.  Leopold also 
loved classical music and German poetry, and enjoyed taking part in the
activities of the German-speaking community in Tallahassee.

All of these aspects of Leopold's life can be seen as reflections of
his own love of beauty and harmony, and that extends also to his practice
as a physicist.  In that regard, it is fitting to close with his own words
from a tribute to the teacher who influenced him most\cite{leo85a}:
``Dirac's greatest impact on me during my ten years associated with him as
a physicist came ... from his conviction that a good theory has to be
beautiful.  This view helped me liberate myself from the bounds of fashion
and made me recognize theoretical physics as an art somewhat akin to music,
capable of expressing ideas that are only vaguely perceived in the back of
one's mind, with the mathematical techniques as the instruments.''

\section*{Acknowledgments}

Leopold, who never married, is survived by a niece, Alisa Lustig of Toronto,
to whom we are indebted for remarks and reminiscences, along with
friends and colleagues including Ron Adler, Francis Everitt,
Bob Jantzen, Kate Kirwin, Hendrik Monkhorst, Zbigniew Oziewicz,
Gabriele Plendl, Merwin Rosenberg, Remo Ruffini, Alex Silbergleit,
Margot Tuckner and Alka Velenik.

\section*{Note about the References}

The following papers and preprints, together with Leopold's correspondence
and other documents, will be available to researchers at the archives
of the Austrian Academy of Sciences at
\url{http://www.oeaw.ac.at/biblio/Archiv/Archiv.html}.

\end{document}